\documentclass[prc,preprint,superscriptaddress,showpacs,nofootinbib,%
tightenlines]{revtex4}
\begin{document}
\preprint{MKPH-T-04-05}
\title{Consistency of the effective-field-theory approach to the
nucleon-nucleon interaction problem revisited}
\author{Jambul Gegelia}
\affiliation{Institut f\"ur Kernphysik, Johannes
Gutenberg-Universit\"at, D-55099 Mainz, Germany} \affiliation{High
Energy Physics Institute, Tbilisi State University, University
St.~9, 380086 Tbilisi, Georgia}
\author{Stefan Scherer}
\affiliation{Institut f\"ur Kernphysik, Johannes
Gutenberg-Universit\"at, D-55099 Mainz, Germany}
\begin{abstract}
It is argued that Weinberg's approach to the nucleon-nucleon (NN) interaction
problem within effective field theory provides a consistent power counting for
renormalized diagrams.
   Within this scheme the NN potential is organized as an
expansion in terms of small quantities like small external momenta and the pion
mass (divided by the characteristic large scale of the effective theory).
   Physical observables to any given order in these small quantities
are calculated from the solutions of the Lippmann-Schwinger (or Schr\" odinger)
equation.
\end{abstract}
\date{March 17, 2004}
\maketitle

\section{\label{introduction} Introduction}

   Weinberg's work on constructing the nuclear forces from effective field theory
(EFT) of the strong interactions \cite{Weinberg:1978kz,Weinberg:rz,Weinberg:um}
has triggered an intensive research activity during the last decade.
   According to Weinberg's approach, in order to calculate any amplitude
involving low-energy pions and nucleons, one first writes down the most general
Lagrangian, draws all Feynman diagrams which contribute to the given process, and
counts powers of small quantities (like small external momenta and the pion mass)
assigned to these diagrams.
   For processes involving $N>1$ nucleons, at any given order one finds an
infinite number of diagrams.
   Weinberg observed that among these diagrams there is only a finite number
of $N$-nucleon irreducible diagrams.
   Defining the sum of these irreducible diagrams as the potential,
he suggested a systematic expansion by applying the power counting to the
potential.
   Next, it is assumed that the renormalized coupling constants are natural, i.e.,
if a coupling constant $C$ has dimension $(mass)^{-d}$, then $\tilde C=C
Q^{d}\sim 1$, where $Q$ is the characteristic large scale of the EFT.
   Within this assumption the higher-order terms of the
potential are suppressed by powers of small momenta or the pion
mass divided by $Q$.
  The contributions of reducible diagrams are taken into account by
solving the Lippmann-Schwinger (LS) or Schr\"odinger equation.

   The application of these ideas in practical calculations has encountered
various problems.
   They originate from the fact that the approach outlined in
Refs.~\cite{Weinberg:rz,Weinberg:um} does not exactly specify how to handle the
problem of renormalization for the LS equation with nonrenormalizable potentials,
i.e., when the iteration of the potential generates divergent terms with
structures, which are not included in the original potential.

For definiteness let us consider the NN scattering problem, which
has attracted much attention during the last few years (see, e.g.,
Refs.\
\cite{Ordonez:1992xp,Ordonez:1995rz,Kaplan:1996xu,Lepage:1997cs,
Kaiser:1997mw,Richardson:1997dz,Phillips:1997tn,
Kaplan:1998tg,Kaplan:1998we,
Lutz:1998uz,Savage:1998vh,Birse:1998dk,Gegelia:1998ee,Gegelia:gn,
Gegelia:1998xr,Birse:1998tm,Epelbaum:1998ka,Epelbaum:1998hg,Epelbaum:1998na,Epelbaum:mq,
Gegelia:1998iu,Park:1998cu,Lepage:1999kt,
Gegelia:1999ja,Cohen:1999ia,vanKolck:1999mw,Epelbaum:1999dj,
Kong:1999sf,Kaplan:1999qa,Fleming:1999ee,Beane:2000fx,Beane:2001bc,
Gegelia:2001ev,Eiras:2001hu,Bedaque:2002mn,Barford:2002je,Nieves:2003uu,
PavonValderrama:2003np,Birse:2003nz,Entem:2003ft,
Yang:2003kn,Gegelia:2003ta,Cohen:2004kf}).
   The case of several nucleons is analogous.
   Since the NN potential of the EFT is nonrenormalizable in the traditional
sense, for the renormalization of the solution of the LS equation one needs to
take into account the contributions of an infinite number of higher-order
counterterms, where the infinite number refers to both the loop and the chiral
expansions.
   The freedom of choosing the finite parts of these counterterms is compensated
by the running of the corresponding renormalized couplings.
   It has been argued that the coefficients of the divergent parts of the
counterterms contributing in low-order calculations would set the scale of the
renormalized couplings.
   As a consequence, even if these couplings were natural at some value
of the renormalization scale, they would become unnaturally large for slightly
different values of this parameter.
   This problem, in different variations, has been addressed as the inconsistency
of Weinberg's approach,  and alternative power counting schemes have been
suggested (see, e.g., Refs.\
\cite{Kaplan:1998tg,Kaplan:1998we,Savage:1998vh,Beane:2001bc,Bedaque:2002mn}).

   In principle, the parameters of the EFT are determined in terms of QCD.
   The numerical values of the couplings within a given renormalization scheme as
well as their renormalization-group (RG) behavior are uniquely fixed.
   In practice, we are still far from obtaining the values of the coupling constants from
QCD and we are unaware of their true RG behavior, but we keep in mind that, in
principle, they do exist.
   In actual calculations one can use different power counting schemes
and calculate the RG behavior of the coupling constants.
   Given two power counting schemes and
provided that the same renormalization condition has been used,
the difference between the two (from the point of view of the RG
behavior of the coupling constants) can only be that one
approximates the true RG behavior better than the other.
   There cannot be a fundamental problem in one
scheme (like the coupling constants blowing up for a small change of
renormalization scale) while it is absent in the other.
   It is only possible that the RG behavior analysis can be trusted in one scheme
and is unreliable in the other because, say, the beta function is calculated
perturbatively while the perturbation theory diverges badly.
   To be more specific, if the renormalized couplings are well-behaved within
the KSW power counting \cite{Kaplan:1998tg} or the new power counting of
Ref.~\cite{Beane:2001bc}, then they are also well-behaved in Weinberg's power
counting.
   As mentioned above, the actual RG behavior of the renormalized coupling constants
has nothing to do with power counting.

   To the best of our knowledge the RG behavior of the coupling constants
has never been calculated analytically within Weinberg's scheme.
   In practice, the couplings are fitted to experimental data.
   The reasonable success of Weinberg's
approach in describing the experimental data
\cite{Ordonez:1995rz,Epelbaum:1999dj,Entem:2003ft} suggests that
the coupling constants should be well-behaved.
   As will be demonstrated below, the estimates of the contributions into
beta functions also suggest that there is no reason to expect that the
renormalized couplings become unnaturally large.

   The aim of the present paper is to argue that Weinberg's approach can be
rigorously completed (or rather worked out in details) in such a way that it
provides a systematic power counting for renormalized diagrams, allowing one to
calculate physical observables to any given order in small quantities.
   For similar discussions, see also
Refs.~\cite{Lepage:1997cs,Gegelia:1998iu,Gegelia:1998ee,Park:1998cu,Lepage:1999kt,Gegelia:2001ev}.

\section{\label{gen_con}General considerations about EFT}

   The motivation for doing EFT of the strong
interactions is the understanding that on the one hand there exists a
``fundamental'' theory of the strong interactions, namely QCD,\footnote{QCD
itself is almost certainly a leading-order approximation of some effective theory
\cite{Weinberg:1996kw}.} which on the other hand cannot yet be directly applied
in the low-energy region.
   Let us recall the central features which comprise a consistent approach to
EFT calculations:
\begin{enumerate}
\item Write down the most general possible Lagrangian, consistent with
assumed symmetry principles \cite{Weinberg:1978kz,Weinberg:rz,Weinberg:um}.
\item Consider {\it all} Feynman diagrams which contribute to the
process in question.\footnote{Note that, strictly speaking, one needs to
analyze an infinite number of diagrams to decide which of them contribute
to any given order of perturbation theory.}
\item Since loop diagrams diverge, they are renormalized by absorbing
the infinite parts into the redefinition of the fields and parameters of the
most general Lagrangian.
   According to Refs.\ \cite{Weinberg:1978kz,Weinberg:rz,Weinberg:um} the
renormalization points should be chosen of the order of small external momenta
(but for logarithmic divergences this is not strictly necessary).
   The practical way of performing the renormalization systematically is
to rewrite the Lagrangian in terms of renormalized fields and parameters.
   The expansion of bare quantities in terms of renormalized coupling
constants then produces the main interactions and counter\-terms.
\item Define the power counting: each {\em renormalized} diagram is assigned
a definite power of small quantities such as small external momenta or the pion
mass.
   For simplicity we use the notion ``renormalized diagram'' for the sum
of the unrenormalized value of the basic graph and the sum of the counterterm
graphs.
\item   The existence of a consistent power counting depends on the applied
renormalization scheme.\footnote{
   For example, even in the purely mesonic sector of chiral perturbation theory
using, say, a cutoff regularization with unsuitable renormalization condition
would destroy the traditional consistent power counting formulated within
dimensional regularization accompanied by the $\overline{\rm MS}$ scheme
\cite{Gasser:1997rx}.}
\item   If the renormalized coupling constants are natural for a given
renormalization condition then the higher orders of small parameters are
suppressed.
   In order to calculate a physical quantity to some specified order, identify all {\it
renormalized} diagrams up to the given order and sum them up.\footnote{In this
paper it is understood that up to some order stands for up to and including this
order.}
   The result is, in general, renormalization-scheme independent only up to the
given order.
   This general feature of quantum field theories is also characteristic for EFT.
\item For processes with more than one nucleon one finds an infinite number of diagrams
at any given order. To sum them up one writes down equations for
regularized diagrams (Lippmann-Schwinger equation for NN, Faddeev
equations for NNN, etc.) and renormalizes the solution so that its
perturbative expansion reproduces the sum of an infinite number of
renormalized diagrams.
    When solving equations one usually also sums up some sets of higher-order
contributions, but that should not affect the accuracy of the result provided
that the power counting is at work, i.e., the contributions of higher-order
diagrams are indeed suppressed.

   In most cases it is technically impossible to solve the regularized
equations analytically and thus one is not able to remove all divergences
explicitly.
   Rather, one solves the equations for the regularized expressions
numerically, matches the low-energy coupling constants to experimental data,
and keeps the regularization parameter finite.
   This procedure is as reliable as the subtractive renormalization, if
it is possible to choose the regularization parameter so that the difference
between the properly renormalized (subtracting every single divergence and
removing the regularization) and the fitted (with finite regulator) expressions
of physical quantities is of order higher than the given accuracy.
   This is in fact the case provided that the renormalized couplings are
natural and the cutoff parameter is kept of the order of the characteristic large
scale of the theory.\footnote{An estimate for this scale is provided by the mass
of the lightest particle which has been integrated out.}

\item Although the EFT contains an infinite number of renormalized low-energy
coupling constants, in principle, they are fixed by QCD.
   In practice, at any given order only a finite number of parameters
contribute, which can be estimated using experimental information.
   The so determined values can be used in calculations of other quantities,
i.e., one can make predictions at the given order.
\end{enumerate}

\section{\label{section:nneft}The NN scattering problem in EFT}
   Once the effective Lagrangian is known, a calculation of the NN scattering amplitude
within EFT requires to specify the regularization.
   It is preferable to use a
regularization which preserves the symmetries of the theory (for a
chiral-symmetry-preserving  ``cutoff'' regularization of the nonlinear sigma
model, see \cite{Slavnov:aw}).\footnote{Unfortunately,  dimensional
regularization does not seem to be useful when working with equations except
for some simple cases, which can be solved exactly.}
    Otherwise one needs to take special care of symmetry-breaking
effects generated by the regularization.
    We will assume that some kind of regularization has been introduced and
proceed without specifying its exact form, using the term ``cutoff'' for the
regularization parameter.
   Next one rewrites the Lagrangian in terms of renormalized quantities.
   The nucleon-nucleon scattering amplitude up to order $n$ in
small expansion parameters is, in principle, obtained by summing up an infinite
number of renormalized diagrams in the framework of old-fashioned
(time-ordered) perturbation theory.
   This can, at least formally, be done by substituting the following
potential (parameterized in terms of renormalized couplings) in the
Lippmann-Schwinger (LS) equation\footnote{The solution of the LS equation will
also contain the contributions of an infinite number of {\it higher-order}
renormalized diagrams.}
\begin{equation}
V^{(n)}\left(E, {\bf p'}, {\bf p}\right)=\sum_{i=0}^\infty \hbar^i
\sum_{j=0}^{(i+1)n+i} C^{(j)}_i \left( {\bf p'}, {\bf p}\right)+ \sum_{i=0}^m
\hbar^i \sum_{j=0}^{n} V^{(j)}_i \left(E, {\bf p'}, {\bf p}\right),
\label{expofV}
\end{equation}
where we have displayed both its chiral as well as loop expansions (expansion in
$\hbar$).
   The $C$ terms stand for NN contact interactions and the
$V$ terms for all other contributions.
   The chiral expansion of the $C$ part of the NN potential contains an infinite number
of terms, because the EFT is not renormalizable in the traditional sense.
   At any fixed order in the loop expansion, the chiral expansion contains
a finite number of terms.
   This number can be calculated by counting the so-called overall (or superficial)
degree of divergence of the loop diagrams which are generated by iterating
the LS equation, and demanding that all these divergences
are cancelled by counterterm contributions.
   The $V$ part of the potential contains a finite number of terms,
i.e.~$m=m(n)$ is a finite number for any finite $n$ and its functional form
is determined by the power counting.
   All terms in  Eq.~(\ref{expofV}) having chiral orders larger than $n$ are
contributions of counterterms, generated by the loop expansion of the bare
couplings of the NN contact interaction terms.
   In general, the $V$ part of the potential depends on the
energy $E$,\footnote{One can also consider energy-independent potentials
using a unitary transformation
\cite{Epelbaum:1998ka,Epelbaum:1999dj}.}
while this dependence has been eliminated from the $C$ part using the
equation of motion or a field transformation
\cite{Weinberg:um,Scherer:1994wi}.

   The amplitude $T^{(n)}\left(E, {\bf p'}, {\bf p}\right)$ is obtained by
solving the LS equation
\begin{equation}
T^{(n)}\left(E, {\bf p'}, {\bf p}\right)=V^{(n)}\left(E, {\bf p'},
{\bf p}\right)+\int \frac{d^3{\bf q}}{(2\pi)^3} \ V^{(n)}\left(E,
{\bf p'}, {\bf q}\right) G(E,{\bf q})\ T^{(n)}\left(E, {\bf q}, {\bf
p}\right), \label{lseq}
\end{equation}
where $G$ is the free two-nucleon propagator.
   It is understood that the loop integration in the LS equation
is regularized.
   If the regularization is introduced on the level of the EFT
Lagrangian, then the LS equation is automatically regularized.
   The counterterms of the Lagrangian are fixed by demanding that
the divergences originating from both the loop diagrams of the potential
and the iteration of the LS equation are cancelled.
   The finite parts of the counterterms are fixed by the choice of
the renormalization condition (see item 3.~of Sec.~\ref{gen_con}).
   The net result of loop diagrams and corresponding counterterms
depends on the chosen renormalization condition.
   This renormalization-scheme dependence is
exactly compensated by the running of the renormalized couplings so that the
scattering amplitudes remain renormalization-scheme independent up to the given
order of accuracy.

   In practice it is not feasible to specify the infinite
number of terms contributing to the potential $V^{(n)}$ of Eq.~(\ref{expofV}).
   However, as we will argue below, this is not necessary.
   Instead of the potential of Eq.~(\ref{expofV}) one rather considers
\begin{eqnarray}
{\cal V}^{(n)}\left(E, {\bf p'}, {\bf p}\right)&=&
\sum_{j=0}^{n} c^{(j)} \left( {\bf p'},{\bf p}\right)
+\sum_{j=0}^{n} v^{(j)} \left(E, {\bf p'}, {\bf p}\right)\nonumber\\
&\equiv&\sum_{j=0}^{n} \sum_{i=0}^\infty \hbar^i c^{(j)}_i
\left( {\bf p'}, {\bf p}\right)
+\sum_{j=0}^{n} \sum_{i=0}^m \hbar^i  V^{(j)}_i \left(E, {\bf p'}, {\bf
p}\right).
\label{expofVpr}
\end{eqnarray}
   The second (non-contact-interaction) part coincides with the second part of
Eq.~(\ref{expofV}) but the first part (purely contact-interaction terms) contains
only terms up to order $n$ in the chiral expansion.
   Note that, in general, $c^{(n)}_i \left( {\bf p'},{\bf p}\right)
\neq C^{(n)}_i \left( {\bf p'},{\bf p}\right)$ for $i\neq 0$.
   One substitutes $v^{(n)}$ into the LS equation, solves numerically, and fits
the available free parameters to data.

   Let us have a closer look at the implications of such an approach.
   The counterterm contributions in Eq.~(\ref{expofVpr}) are fixed so that
all divergences generated by iterations of the LS equation with
${\cal V}^{(n)}$, the coefficients of which are of chiral order up
to $n$, are cancelled.
   The remaining divergent parts are of order $n+1$ or higher
in the chiral expansion.
   Moreover, all cutoff-dependent terms of order $n$ or less
which vanish in the removed-regularization limit (i.e. depend on
inverse powers of the cutoff) can also be removed using the
$c^{(j)}_i$.
    The difference between the $C^{(j)}_i$ and $c^{(j)}_i$ of
Eqs.\ (\ref{expofV}) and (\ref{expofVpr}), respectively, occurs due
to the fact that the $c^{(j)}_i$ contain terms which absorb the inverse
powers of the cutoff and, more importantly, the $C^{(j)}_i$ contain
additional terms which absorb divergences generated by iterating higher-order
counterterms.
   Choosing the counterterm contributions as specified above, one obtains the
amplitude which, to order $n$, coincides term by term with the amplitude
which was obtained by iterating the potential of Eq.~(\ref{expofV}).
   The difference between the two results depends on the cutoff parameter
$\Lambda$ and contains contributions like
\begin{equation}
\sim \left( \frac{q^2}{\Lambda^2}\right)^{i} \ \ {\rm with} \ \ i>
n/2 \ \ {\rm as \ well \ as} \ \ \sim \frac{\left(
q^2\right)^{i}\Lambda^j}{Q^{2i+j}} \  {\rm with} \ \ i> n/2, \ \
j>0, \label{divterms}
\end{equation}
where $q$ denotes a small external momentum or the pion mass and $Q$ the
characteristic large scale of the EFT.
   To keep both types of contributions suppressed, one has to choose
$\Lambda\sim Q$.
   For an example in the context of contact interactions plus a
one-pion-exchange potential in the ${ }^1S_0$ channel,
see Ref.~\cite{Gegelia:2001ev}.

   In subtractively renormalized EFT the contributions of an infinite number
of counter\-terms of Eq.~(\ref{expofVpr}) should have been taken into account so
that all positive powers of the cutoff are cancelled.
   Then one should and also could consider the removed regularization
limit.\footnote{To practically realize this scheme (in numerical calculations)
one could use Wilson's renormalization-group approach \cite{Wilson:1973jj}.
   For an application to NN-potential models, see
Refs.\ \cite{Bogner:2001gq,Bogner:2001jn,Bogner:2002yw,Bogner:2003wn}.}
   On the other hand, if one substitutes Eq.~(\ref{expofVpr}) in the LS
equation and fits the available parameters to physical quantities, then one
approximates the rigorous result of subtractively renormalized EFT up to the
given accuracy {\it only if one takes $\Lambda\sim Q$}.
   Taking instead $\Lambda\to\infty$, the second type of contributions of
Eq.~(\ref{divterms}) will dominate the amplitude instead of being
discarded (i.e.~subtracted),
because the corresponding counterterms are not included.
   Therefore, although it might be interesting to consider such kind of
{\em models} they have little to do with EFT.
   In this context, one can encounter problems which actually are not
related with EFT, like being unable to describe a positive effective range or
different regularizations leading to different results.

   It is worth noting that by taking the potential of Eq.~(\ref{expofVpr}),
solving the corresponding LS equation, fitting the
parameters to some set of physical quantities,
and substituting these parameters in any remaining physical quantity,
one automatically performs the absorption of the corresponding divergences.
   One should therefore keep in mind that what is fitted to physical
quantities are not the renormalized couplings of $c_0^{(j)}$ but rather the bare
couplings of $c^{(j)}$ in Eq.~(\ref{expofVpr}).
   On the other hand, the couplings which contribute
in the second (non-contact-interaction) part of the potential are the
renormalized couplings.

   One could argue that the above discussion, although applied to an infinite
number of diagrams, is still perturbative.
   There might be nonperturbative contributions to the solutions of
the LS equation which have trivial perturbative expansions.
   While this is certainly possible, the EFT provides a systematic
power counting only for those parts of the solutions which,
if expanded, reproduce perturbative diagrams term by term.
   One should handle the nonperturbative EFT problem in a way which is
consistent with perturbative expansions.
   While this is a necessary condition we are not in a position to argue
whether or not it is, in general, also sufficient.

\section{\label{section:estimate}Estimating renormalized coupling constants}
   In this section we will estimate some contributions to the running of
the renormalized couplings within Weinberg's approach which have been
pointed out as a possible source of the problem of unnaturally large couplings
\cite{Kaplan:1996xu,Savage:1998vh}.
   To address the issue of the renormalization-group behavior of the running
couplings let us consider the leading-order potential
\begin{equation}
V_0\left( {\bf p'},{\bf p}\right)=C-\left( \frac{g_A^2}{4 F^2}\right) \
\frac{\left( {\bf q}\cdot\sigma_1 \ {\bf q}\cdot\sigma_2\right)\left(
\tau_1\cdot\tau_2\right)} {{\bf q^2}+M_\pi^2}, \label{potential}
\end{equation}
where ${\bf q}={\bf p'}-{\bf p}$, $C$ stands for a contact-interaction
contribution in a particular spin-isospin channel, $g_A=1.267$ is the
axial-vector coupling constant, $F=92.4$ MeV the pion decay constant, and $M_\pi=
139.6$ {\rm MeV} the charged-pion mass.
   Substituting $V_0$ into the LS equation and iterating,
(divergent) loop diagrams are generated.
   The divergences have to be absorbed by contributions of the counterterms.
   The coefficients of the divergent parts of the counterterms are closely
related to the running behavior of the corresponding renormalized couplings.

   Let us consider an example which has previously been used to
demonstrate a problem with the renormalization-group behavior of the
renormalized couplings \cite{Kaplan:1996xu,Savage:1998vh}.
   We substitute the potential of Eq.~(\ref{potential}) for the
$^1S_0$ channel into the LS equation, iterate twice, and consider the
contribution proportional to $C^2 {g_A^2}/{4 F^2}$.
   In dimensional regularization its divergent part reads \cite{Savage:1998vh}
\begin{equation}
-\frac{1}{\epsilon}\frac{g_A^2 M_\pi^2 m^2_N}{256 \pi^2 F^2} \ C^2,
\label{firstcountterm}
\end{equation}
where $m_N$ is the nucleon mass.
   This divergence has to be cancelled by a contribution of a counterterm
generated by the loop expansion of a $D_0 M_\pi^2$ contact interaction term,
where $D_0$ is the bare coupling.
   Equation (\ref{firstcountterm}) leads to following running of the
renormalized coupling
\begin{equation}
D (\mu)=D (\mu_0 )+\frac{g_A^2 m^2_N}{256 \pi^2 F^2} \
C^2 \ln\left(\frac{\mu}{\mu_0}\right).
\label{Drunning}
\end{equation}
   If we take $D\left( \mu_0\right)$ negligible and
$\ln(\mu/\mu_0)\sim 1$, then $D$ is dominated by second term in
Eq.~(\ref{Drunning}).
   Comparing the $D M_\pi^2$ term with the leading-order contact-interaction
term, we obtain
\begin{equation}
\frac{D M_\pi^2}{C}=\frac{g_A^2 m^2_N M_\pi^2}{256 \pi^2 F^2} \ C\approx C \
\frac{M_\pi^2}{15}\approx \frac{M_\pi^2}{(430 \ {\rm MeV})^2}, \label{dvc}
\end{equation}
where we made use of the estimate of $C\approx 1/(110 \ {\rm MeV})^2$ of
Ref.~\cite{Kaplan:1998tg}.
   Equation (\ref{dvc}) suggests that the coupling $D$ does not get
enhanced and there is no need to promote it to the leading order.

   Next we consider the ladder diagrams, obtained by iterating the
one-pion-exchange potential.
   To be specific, let us take the $2n$ loop diagram contributing
to the ${ }^3S_1-{ }^3D_1$ channel NN scattering.
   One can estimate the coefficient of the logarithmically divergent part of
the $2n$-loop diagram as
\begin{equation}
\sim \frac{g_A^2}{4 F^2} \ \left( \frac{1}{(4 \pi )^{3/2}}
\frac{g_A^2}{4 F^2} m_N\right)^{2 n} (q^2)^n\approx
\frac{1}{\left(146 \ {\rm MeV}\right)^2} \ \left(
\frac{q^2}{(1010\ {\rm MeV})^2}\right)^n, \label{loopfactors}
\end{equation}
where $q$ stands for a small momentum.
   The contribution to the renormalized coupling of the
$\left(q^2\right)^n$ term corresponding to Eq.~(\ref{loopfactors})
for $\ln(\mu/\mu_0)\sim 1$  is $$ \sim \frac{1}{\left(146 \ {\rm
MeV}\right)^2} \ \left( \frac{1}{1010 \ {\rm MeV}}\right)^{2n}. $$
   From the above estimates we cannot conclude that the
renormalized couplings of higher-order contact interactions would become
unnaturally large due to the fact that the corresponding counterterms
contribute to the renormalization of loop diagrams of lower order
in the chiral expansion.

\section{Conclusions}

   We have argued that Weinberg's approach to nuclear physics problems
in the framework of EFT is free of conceptual inconsistencies, once the power
counting is applied to renormalized diagrams as opposed to unrenormalized
diagrams and counterterm contributions separately.
   In order to solve the regularized equations and to subsequently renormalize
the solutions, in general, it is necessary to take the contributions of an
infinite number of counterterms into account.
   Although, except for a few simple cases, it seems to be impossible to
carry out this program, in practical calculations one does not actually need to
do so.
   For example, for NN scattering one solves the regularized LS equation and fits
the parameters of the effective Lagrangian (up to the given order) to physical
quantities.
   The difference between the so obtained amplitudes and
the exactly renormalized amplitudes is of higher order, provided the renormalized
couplings are natural and the cutoff parameter is chosen of the order of the
characteristic large scale of the EFT.

\acknowledgments
   J.~Gegelia acknowledges the support of the Alexander von Humboldt
Foundation and the Deutsche Forschungsgemeinschaft (SFB 443).
   He would also like to thank the Institute for Nuclear Theory at the
University of Washington for its hospitality and financial support.


\begin{thebibliography}{100}

\bibitem{Weinberg:1978kz}
S.~Weinberg,
Physica A {\bf 96}, 327 (1979).


\bibitem{Weinberg:rz}
S.~Weinberg,
Phys.\ Lett.\ B {\bf 251}, 288 (1990).

\bibitem{Weinberg:um}
S.~Weinberg,
Nucl.\ Phys.\ {\bf B363}, 3 (1991).

\bibitem{Ordonez:1992xp}
C.~Ordonez and U.~van Kolck,
Phys.\ Lett.\ B {\bf 291}, 459 (1992).

\bibitem{Ordonez:1995rz}
C.~Ordonez, L.~Ray, and U.~van Kolck,
Phys.\ Rev.\ C {\bf 53}, 2086 (1996).

\bibitem{Kaplan:1996xu}
D.~B.~Kaplan, M.~J.~Savage, and M.~B.~Wise,
Nucl.\ Phys.\ {\bf B478}, 629 (1996).

\bibitem{Lepage:1997cs}
G.~P.~Lepage,
arXiv:nucl-th/9706029.

\bibitem{Kaiser:1997mw}
N.~Kaiser, R.~Brockmann, and W.~Weise,
Nucl.\ Phys.\ {\bf A625}, 758 (1997).

\bibitem{Richardson:1997dz}
K.~G.~Richardson, M.~C.~Birse, and J.~A.~McGovern,
arXiv:hep-ph/9708435.

\bibitem{Phillips:1997tn}
D.~R.~Phillips, S.~R.~Beane, and T.~D.~Cohen,
Nucl.\ Phys.\ {\bf A631}, 447c (1998).

\bibitem{Kaplan:1998tg}
D.~B.~Kaplan, M.~J.~Savage, and M.~B.~Wise,
Phys.\ Lett.\ B {\bf 424}, 390 (1998).

\bibitem{Kaplan:1998we}
D.~B.~Kaplan, M.~J.~Savage, and M.~B.~Wise,
Nucl.\ Phys.\ {\bf B534}, 329 (1998).


\bibitem{Lutz:1998uz}
M.~Lutz,
Nucl.\ Phys.\ {\bf A642}, 171 (1998).

\bibitem{Savage:1998vh}
M.~J.~Savage,
arXiv:nucl-th/9804034.

\bibitem{Birse:1998dk}
M.~C.~Birse, J.~A.~McGovern, and K.~G.~Richardson,
Phys.\ Lett.\ B {\bf 464}, 169 (1999).


\bibitem{Gegelia:1998ee}
J.~Gegelia,
arXiv:nucl-th/9806028.

\bibitem{Gegelia:gn}
J.~Gegelia,
Phys.\ Lett.\ B {\bf 429}, 227 (1998).

\bibitem{Gegelia:1998xr}
J.~Gegelia,
arXiv:nucl-th/9802038.

\bibitem{Birse:1998tm}
M.~C.~Birse, J.~A.~McGovern and K.~G.~Richardson,
arXiv:hep-ph/9808398.

\bibitem{Epelbaum:1998ka}
E.~Epelbaum, W.~Gl{\"o}ckle, and U.~-G.~Mei\ss ner,
Nucl.\ Phys.\  {\bf A637}, 107 (1998).

\bibitem{Epelbaum:1998hg}
E.~Epelbaum, W.~Gl{\"o}ckle, and U.~-G.~Mei\ss ner,
Phys.\ Lett.\ B {\bf 439}, 1 (1998).

\bibitem{Epelbaum:1998na}
E.~Epelbaum, W.~Gl{\"o}ckle, A.~Kruger, and U.~-G.~Mei\ss ner,
Nucl.\ Phys.\ {\bf A645}, 413 (1999).

\bibitem{Epelbaum:mq}
E.~Epelbaum, W.~Gl{\"o}ckle, and U.~-G.~Mei\ss ner,
Few Body Syst.\ Suppl.\  {\bf 10}, 479 (1999).

\bibitem{Gegelia:1998iu}
J.~Gegelia,
J.\ Phys.\ G {\bf 25}, 1681 (1999).

\bibitem{Park:1998cu}
T.~S.~Park, K.~Kubodera, D.~P.~Min, and M.~Rho,
Nucl.\ Phys.\ {\bf A646}, 83 (1999).

\bibitem{Lepage:1999kt}
G.~P.~Lepage, {\it Conference summary,}
{Prepared for INT Workshop on Nuclear Physics with Effective Field
Theory, Seattle, Washington, 25-26 Feb 1999.}

\bibitem{Gegelia:1999ja}
J.~Gegelia,
Phys.\ Lett.\ B {\bf 463}, 133 (1999).

\bibitem{Cohen:1999ia}
T.~D.~Cohen and J.~M.~Hansen,
Phys.\ Rev.\ C {\bf 59}, 3047 (1999).

\bibitem{vanKolck:1999mw}
U.~van Kolck,
Prog.\ Part.\ Nucl.\ Phys.\  {\bf 43}, 337 (1999).


\bibitem{Epelbaum:1999dj}
E.~Epelbaum, W.~Gl{\"o}ckle, and U.~-G.~Mei\ss ner,
Nucl.\ Phys.\ {\bf A671}, 295 (2000).

\bibitem{Kong:1999sf}
X.~Kong and F.~Ravndal,
Nucl.\ Phys.\ {\bf A665}, 137 (2000).

\bibitem{Kaplan:1999qa}
D.~B.~Kaplan and J.~V.~Steele,
Phys.\ Rev.\ C {\bf 60}, 064002 (1999).

\bibitem{Fleming:1999ee}
S.~Fleming, T.~Mehen, and I.~W.~Stewart,
Nucl.\ Phys.\ {\bf A677}, 313 (2000).

\bibitem{Beane:2000fx}
S.~R.~Beane, P.~F.~Bedaque, W.~C.~Haxton, D.~R.~Phillips and
M.~J.~Savage,
arXiv:nucl-th/0008064.

\bibitem{Beane:2001bc}
S.~R.~Beane, P.~F.~Bedaque, M.~J.~Savage, and U.~van Kolck,
Nucl.\ Phys.\ {\bf A700}, 377 (2002).


\bibitem{Gegelia:2001ev}
J.~Gegelia and G.~Japaridze,
Phys.\ Lett.\ B {\bf 517}, 476 (2001).

\bibitem{Eiras:2001hu}
D.~Eiras and J.~Soto,
Eur.\ Phys.\ J.\ A {\bf 17}, 89 (2003).

\bibitem{Bedaque:2002mn}
P.~F.~Bedaque and U.~van Kolck,
Ann.\ Rev.\ Nucl.\ Part.\ Sci.\  {\bf 52}, 339 (2002).

\bibitem{Barford:2002je}
T.~Barford and M.~C.~Birse,
Phys.\ Rev.\ C {\bf 67}, 064006 (2003).

\bibitem{Nieves:2003uu}
J.~Nieves,
Phys.\ Lett.\ B {\bf 568}, 109 (2003).

\bibitem{PavonValderrama:2003np}
M.~Pavon Valderrama and E.~Ruiz Arriola,
Phys.\ Lett.\ B {\bf 580}, 149 (2004).

\bibitem{Birse:2003nz}
M.~C.~Birse and J.~A.~McGovern,
arXiv:nucl-th/0307050.

\bibitem{Entem:2003ft}
D.~R.~Entem and R.~Machleidt,
Phys.\ Rev.\ C {\bf 68}, 041001(R) (2003).

\bibitem{Yang:2003kn}
J.~F.~Yang,
arXiv:nucl-th/0310048.

\bibitem{Gegelia:2003ta}
J.~Gegelia,
Eur.\ Phys.\ J.\ A {\bf 19}, 355 (2004).

\bibitem{Cohen:2004kf}
T.~D.~Cohen, B.~A.~Gelman, and U.~van Kolck,
arXiv:nucl-th/0402054.

\bibitem{Weinberg:1996kw}
S.~Weinberg, arXiv:hep-th/9702027.

\bibitem{Gasser:1997rx}
J.~Gasser,
arXiv:hep-ph/9711503.

\bibitem{Slavnov:aw}
A.~A.~Slavnov,
Nucl.\ Phys.\ {\bf B31}, 301 (1971).

\bibitem{Scherer:1994wi}
S.~Scherer and H.~W.~Fearing,
Phys.\ Rev.\ D {\bf 52}, 6445 (1995).

\bibitem{Wilson:1973jj}
K.~G.~Wilson and J.~B.~Kogut,
Phys.\ Rept.\  {\bf 12}, 75 (1974).

\bibitem{Bogner:2001gq}
S.~K.~Bogner, T.~T.~S.~Kuo, A.~Schwenk, D.~R.~Entem, and
R.~Machleidt,
Phys.\ Lett.\ B {\bf 576}, 265 (2003).

\bibitem{Bogner:2001jn}
S.~K.~Bogner, A.~Schwenk, T.~T.~S.~Kuo, and G.~E.~Brown,
arXiv:nucl-th/0111042.

\bibitem{Bogner:2002yw}
S.~Bogner, T.~T.~S.~Kuo, L.~Coraggio, A.~Covello, and N.~Itaco,
Phys.\ Rev.\ C {\bf 65}, 051301 (2002).

\bibitem{Bogner:2003wn}
S.~K.~Bogner, T.~T.~S.~Kuo, and A.~Schwenk,
Phys.\ Rept.\  {\bf 386}, 1 (2003).


\end{thebibliography}
\end{document}